\documentclass{IEEEtran}
\usepackage{amsmath,amssymb,amsfonts}
\usepackage{algorithmic}
\usepackage{graphicx}
\usepackage[table]{xcolor}
\usepackage{textcomp}
\usepackage[hidelinks]{hyperref}

\usepackage{multirow}
\usepackage{booktabs}
\newcolumntype{C}{>{$}c<{$}}
\AtBeginDocument{
\heavyrulewidth=.08em
\lightrulewidth=.05em
\cmidrulewidth=.03em
\belowrulesep=.65ex
\belowbottomsep=0pt
\aboverulesep=.4ex
\abovetopsep=0pt
\cmidrulesep=\doublerulesep
\cmidrulekern=.5em
\defaultaddspace=.5em
}

\def\BibTeX{{\rm B\kern-.05em{\sc i\kern-.025em b}\kern-.08em
    T\kern-.1667em\lower.7ex\hbox{E}\kern-.125emX}}
\usepackage[
    backend=biber,
    style=numeric-comp,
    sortcites,
    sorting=none,
    firstinits=true,
    natbib,
    hyperref,
    maxbibnames=99,
    doi=false,isbn=false,url=false,eprint=false
]{biblatex}

\newcommand{\changed}[1]{#1}

\begin{document}
\title{\changed{Interpretable Classification of Bacterial Raman Spectra with Knockoff Wavelets}}
\author{Charmaine Chia$^*$, Matteo Sesia$^*$, Chi-Sing Ho, Stefanie S. Jeffrey, Jennifer Dionne,\\Emmanuel J. Cand{\`e}s, and Roger T. Howe
\thanks{$^*$Equal contribution.}
\thanks{This work was partly supported by NSF grants DMS
1712800 and 1934578, a Math+X grant (Simons
Foundation), and by the Stanford School of
Engineering's Catalyst for Collaborative Solutions. \changed{The authors thank the Research Computing Center at Stanford University and the Center for Advanced Research Computing at the University of Southern California for providing computing resources.}}
\thanks{C. Chia (email: charmaine@aetherbio.com) was in the Department of
Electrical Engineering at Stanford University, Stanford, CA 94305, coadvised
by R. T. Howe and S. S. Jeffrey. She is now with Aether Biomachines, Menlo
Park, CA 94025 USA.
}
\thanks{M. Sesia (email: sesia@marshall.usc.edu) was in the Department of Statistics at
Stanford University, advised by E. Cand{\`e}s. He is now in the Department of Data Sciences and Operations at the University of Southern California, Los Angeles, CA 90089 USA.}
\thanks{C-S. Ho was in the Department of Applied Physics at Stanford University,
advised by J. Dionne. She is now with Tempus Labs, Redwood City, CA 94065, USA.}
}

\maketitle

\begin{abstract}
\changed{Deep neural networks and other sophisticated machine learning models are widely applied to biomedical signal data
because they can detect complex patterns and compute accurate predictions.
However, the difficulty of interpreting such models is a limitation, especially for applications involving high-stakes decision,
including the identification of bacterial infections.
In this paper, we consider fast Raman spectroscopy data and demonstrate that a logistic regression model 
with carefully selected features achieves accuracy comparable to that of neural networks, while being much simpler and more transparent.
Our analysis leverages wavelet features with intuitive chemical interpretations, and performs controlled variable selection with knockoffs to ensure the predictors are relevant and non-redundant.
Although we focus on a particular data set, the proposed approach is broadly applicable to other types of signal data for which interpretability may be important.
}
\end{abstract}


\section{Introduction}
\label{sec:introduction}
\IEEEPARstart{N}{ew} sensor technologies have contributed to the advent of ``big data" in biomedicine, of which signal data are an important modality. From one-dimensional electrocardiography and electroencephalography signals from the heart and brain, to two-dimensional tissue images of tumor histology, to three-dimensional magnetic resonance images, these consist of sequential measures of an observable along one or more independent axes such as time, distance, or frequency. Signal data differ from structured forms of data in that the meaning of each independent variable is not as distinctively and intuitively definable. Informative features must be extracted using signal processing and machine learning (ML) techniques before useful patterns can be detected and leveraged to make predictions.

While predictive accuracy is usually prioritized in ML, model interpretability is gaining more attention. Interpretability is crucial when models inform the decisions of experts and can have serious consequences, such as in applications involving healthcare. Furthermore, when the signal source itself is not well-understood, interpretable models can yield deeper insights and facilitate inferences.
Along these lines, the ML framework discussed in~\cite{murdoch2019definitions} proposes three metrics for evaluating models: 1) predictive accuracy (the goodness-of-fit to the underlying data), 2) descriptive accuracy (the fidelity of the interpretation in describing relations learned by the model), and 3) relevancy (the usefulness and comprehensibility of the interpretation to the target audience). 

Simpler models ({\em e.g.}, linear regression, trees, naive Bayes) are easier to interpret, though often at the expense of predictive accuracy due to their limited flexibility. By contrast, sophisticated models such as deep neural networks \cite{krishnan2018trends,bengio2003neural,ravi2016deep} can automatically extract predictive features and capture complex relations in the data, but their ``black-box'' nature makes it difficult to understand their decisions.
Various techniques have been proposed to improve the descriptive accuracy of ML models; for example, saliency methods help visualize the activation of individual input features \cite{borji2019saliency}, while attribution methods like LIME \cite{ribeiro2016should} and SHAP \cite{lundberg2017unified} quantify the impact of each feature on the output predictions.
However, these {\em post hoc} techniques are \changed{not designed} for developing simpler models.

With regard to relevancy, studies report that people favor explanations that are short, contrast instances with different outcomes, and highlight abnormal causes \cite{miller2019explanation}. In other words, we seek to understand which features are important, and how these affect the outcome. Data scientists often pursue these goals through feature selection, in addition to feature extraction, to ensure that their conclusions are based on relevant and non-redundant predictors. 
For example, one may want to identify a smaller set of genetic variants linked to disease susceptibility among thousands of possibilities \cite{sesia2020}, or to identify which specific morphological features from brain electroencephalogram signals can diagnose epilepsy \cite{wang2014extracting}.

\changed{There exists a broad literature on variable selection methods designed to identify a subset of important and non-redundant predictors from a large set of features; see \cite{saeys2007review,chandrashekar2014survey,Li2017,heinze2018variable} for an overview. However, most existing techniques are either heuristic, in the sense that they lack clear statistical guarantees, or require asymptotic approximations and strong modeling assumptions, which may not be justified when working with complex biomedical data. Consequently, tuning these models through variable selection may be difficult and their output may include unexpected numbers of false discoveries: unimportant features that are either irrelevant or redundant (see Appendix~\ref{app-methods} for a more precise definition of this concept).
  For example, the lasso is a very successful variable selection method for high-dimensional linear models~\cite{tibshirani1996regression} and it is known to be asymptotically consistent under certain assumptions~\cite{zhao2006model,candes2009near}; it tends to select all relevant non-redundant features as the sample size grows. In practice, however, it often utilizes more predictors than necessary.
Feature selection becomes even more challenging when it involves non-parametric models, although some theoretical results have been obtained for random forests~\cite{scornet2015consistency}, and several proposals have been advanced for sparse neural networks \cite{leray1999feature,verikas2002feature,srinivas2017training}.

A general approach to variable selection with a clear statistical interpretation is offered by the knockoff filter; this was first proposed in the context of linear regression~\cite{barber2015controlling} and later extended to general machine learning algorithms~\cite{candes2018}, including the classification ones considered in this paper.
The main idea of this solution is to augment the available features with an equal number of synthetic negative controls (the {\em knockoffs}). Knockoffs are constructed to be statistically indistinguishable from those variables among the original ones that are unimportant~\cite{candes2018}; however, the identities of the knockoffs are known exactly, unlike those of the latter. Therefore, the important features can be selected by looking for those that significantly stand out from the knockoffs~\cite{barber2015controlling}. See Appendix~\ref{app-methods} for a review of this method.
Under relatively mild assumptions, the knockoff filter is guaranteed to control the false discovery rate (FDR)~\cite{benjamini1995}: the expected proportion of irrelevant or redundant features among the selected ones.
Knockoffs can be applied with any machine learning algorithm and require no modelling assumptions about the unknown relation between the available features and the true bacterial classes.
This flexibility makes knockoffs particularly well-suited to our problem because bacterial classification is an inherently complex task with implications for patient treatment, making robustness and interpretability important considerations.
Furthermore, controlling the FDR is a reasonable objective in our context because we seek to construct predictive models that are both accurate, leveraging all relevant information, and simple to explain, avoiding unnecessary features.
}

\changed{Previous applications of knockoffs have focused on structured data}, in which the features are well-defined a priori: single-nucleotide polymorphisms  \cite{katsevich2017multilayer,sesia2019,shen2019false,sesia2020,sesia2020b}, virus mutations \cite{romano2019}, or demographic/behavioral cancer biomarkers \cite{gimenez2019knockoffs}, to name some examples.
Only few extensions to unstructured data have been reported, namely involving computed tomography (CT) \cite{li2018knockoff}, functional magnetic resonance images \cite{nguyen2019}, and economic time series \cite{fan2019ipad}.
Thus, the relatively unexplored area of unstructured data provides an interesting use case.

In this paper, we combine feature extraction and selection to obtain a powerful and interpretable signal analysis \changed{method}, and demonstrate its utility by applying it to a data set of fast Raman spectroscopy measurements of common bacteria collected at the Stanford Hospital \cite{ho2019rapid}.
Raman spectroscopy measures the interaction of laser light with a sample, producing a spectrum where peaks indicate wavelengths at which the light is strongly absorbed by the chemical bonds present therein. This technique thus yields an optical fingerprint of the sample.
Fast Raman measurements follow the same principle, but their spectra are noisier and more difficult to recognize due to shorter measurement times.
Therefore, reliable ML algorithms are useful to automate the recognition of such optical fingerprints.
Recently, a convolutional neural network (CNN) was found to be successful at using these data to 
predict outcomes such as bacterial strain and antibiotic susceptibility \cite{ho2019rapid}.
These results are promising because rapid and culture-free pathogen identification could advance the treatment of bacterial infections and sepsis. 
At the same time, such high-stakes medical decisions call for more interpretable models that can be easily examined and understood by humans who, for instance, may wish to know the presence of which chemical bonds drives the machine decision.

Our approach begins with a feature extraction step that transforms the signal data into a more intuitive representation
summarizing the presence of localized peaks in the spectra. Then, we apply the knockoff filter to select a subset of features that are likely to be predictive and non-redundant, and finally we use these to fit a simple multinomial logistic regression model that predicts the outcome of interest.
Our analysis shows that the proposed method performs similarly to the CNN of \cite{ho2019rapid} in terms of predictive accuracy, and sometimes even better, while creating a more compact and interpretable model.


\section{DATA SET}

We \changed{analyze} data consisting of 60,000 Raman spectra of dried monolayer bacteria and yeast samples taken with fast (one-second) scans, from \cite{ho2019rapid}. Thirty distinct isolates were measured, including multiple isolates of Gram-negative and Gram-positive bacteria, as well as Candida species; 2000 spectra were measured for each isolate, most of which were taken over single cells.
The spectra consist of 992 measurement points evenly distributed in the spectral range of 381.98 to 1792.4 cm$^{-1}$. The measured Raman intensities were normalized to lie between 0 and 1.
Further details about these measurements can be found in \cite{ho2019rapid}. 
The data can be downloaded from \url{https://github.com/csho33/bacteria-ID}. 

In addition to the Raman spectra ($X$), three sets of associated outcome labels $(Y)$ are available from this data set:
\begin{enumerate}
    \item Isolate labels  $\rightarrow$ 30 classes;
    \item Empiric antibiotic treatment $\rightarrow$ 8 classes;
    \item Methicillin resistance of Staphylococcus aureus strains $\rightarrow$ 2 classes
\end{enumerate}
To summarize, the sizes of the data matrices are:
\begin{itemize}
\item Raw signal data ($X$): $60,000 \times 992$;
\item Outcome labels ($Y$): $60,000 \times 1$, except for the 3$^{\text{rd}}$ set of labels, which apply only to Staphylococcus aureus strains, giving a $10,000 \times 1$ outcome matrix.
\end{itemize}
\changed{The code to reproduce our analysis is available from \url{https://github.com/chicanagram/raman-knockoffs}.}

\section{METHODS}

\subsection{Feature extraction}

Feature extraction is the transformation of raw data into a more discriminatory representation for the prediction task. 
There exist a variety of feature extraction methods for signal data, which can be categorized into four broad families \cite{krishnan2018trends}.

\begin{enumerate}
\item Time/position methods extract characteristic properties from specific windows of measurement points.
\item Frequency methods break signals into their spectral components, giving information complementary to the above; {\em e.g.}, the Fourier transform \cite{bracewell1986fourier}.
\item Time/position-frequency methods capture both frequency and time/position information in non-linear and non-stationary signals; {\em e.g.}, the wavelet transform \cite{mallat1999wavelet}.
\item Sparse signal decomposition methods seek sparse data representations in terms of basis sets that are defined empirically; {\em e.g.}, convolutional dictionary learning \cite{garcia2018convolutional}.
\end{enumerate}

In general, different feature extraction methods may be better suited for different kinds of data, and they should be chosen based on their natural interpretability given the dynamics of the signal source or other relevant prior knowledge \cite{krishnan2018trends}. 
In our application, we opt for a discrete wavelet transform (DWT), which projects the signal $X$ onto a compact orthogonal basis set of wave-like oscillations at different frequencies, beginning and ending with zero amplitude.
\changed{The ability of wavelets to capture both frequency and location information is critical to the analysis of Raman spectral data. In fact, the features of natural interest there are the localized peaks indicating the presence of chemical bonds which may distinguish different types of bacteria. Moreover, the DWT provides a compact representation of the signal which can be computed efficiently, unlike the continuous wavelet transform. }
\changed{The basis wavelet we adopt is a 24-point Coiflet with five DWT levels~\cite{beylkin2009fast}. This choice is motivated by the symmetry of coiflets, which facilitates their comparison with the resonance peaks in the spectra clearly visible to the naked eye~\cite{mccreery2005raman}. The filter length (24 points) and number of decomposition levels were chosen to match as well as possible the visible peaks in our spectra. This tuning was carried out manually, visualizing the basis wavelets alongside de-noised signals obtained by averaging fast Raman spectra from multiple bacterial samples from within the same class.}
The result of the transform, $X'$, is a set of 1105 features for each of the 60,000 samples, which represent the concatenated approximation and detail coefficients from the five-level wavelet filtering procedure. 
Starting from the wavelet representation, the original signal can be reconstructed using an Inverse Discrete Wavelet Transform (IDWT). 

\subsection{Knockoff generation}

We generate knockoffs for both the raw data ($X$) and the wavelet features ($X'$) following the model-X method in \cite{candes2018}, as implemented by the second-order knockoff machines in \cite{romano2019}; see Appendix~\ref{app-methods} for relevant technical background.
We apply this algorithm to generate knockoffs that are approximately pairwise
exchangeable with the data in terms of their second moments.
More precisely, we generate the knockoff features $\tilde{X} \in \mathbb{R}^{n \times p}$
given the original features $X \in \mathbb{R}^{n \times p}$ (or, analogously, $X'$) such that the mean vector and
the covariance matrix of $[X, \tilde{X}] \in \mathbb{R}^{n \times 2p}$ match those of $[X, \tilde{X}]_{\text{swap}(j)}$, for any $j \in \{1,\ldots,p\}$. Above, $\text{swap}(j)$ is the operator that swaps $X_j$ with $\tilde{X}_j$. 
Simultaneously, we try to make each element of $\tilde{X}$ as different as possible from the corresponding element of $X$ \cite{candes2018, romano2019}, to maximize the statistical power of the knockoff filter \cite{barber2015controlling}. 
By such construction, the covariance matrices of $[X,\tilde{X}]$ and $[X, \tilde{X}]_{\text{swap}(j)}$, for any $j$, are approximately
\begin{equation}
  G = \begin{bmatrix}
    \Sigma & \Sigma - \text{diag}(s) \\
    \Sigma - \text{diag}(s) & \Sigma
  \end{bmatrix},
\label{eq:covariance}
\end{equation}
where $\Sigma$ is the covariance matrix of $X$ and the vector $s$ is
maximized subject to the constraint that the matrix $G$ be
positive semi-definite \cite{barber2015controlling,candes2018}. 
We refer to \cite{candes2018} and \cite{romano2019} for
further details on knockoff generation.
It is worth mentioning the method in \cite{romano2019} can accommodate a more general construction that also matches higher
moments of $[X, \tilde{X}]$ to those of $[X, \tilde{X}]_{\text{swap}(j)}$, which leads to a
more robust variable selection procedure in some situations, but
seems to make little difference with our data.
Therefore, we focus on second-order knockoffs for simplicity.

\subsection{Feature selection with the knockoff filter}

Following the generation of knockoffs, 
\changed{80\% of the data points are randomly assigned to a training set, and the remaining 20\% are assigned to a test set, which will be utilized only later to evaluate predictive performance, similarly to~\cite{ho2019rapidPreprint}. The augmented raw and wavelet representation training data,} $[X, \tilde{X}]$ and $[X', \tilde{X}']$, are standardized to make their columns have unit variance, and then they are separately provided as input to a classifier, which is trained on each of the three sets of \changed{the corresponding $Y$} labels. 
The number of features available to each model is thus twice that of the original data. 
The classification model is based on logistic regression with $\ell_1$ (lasso) regularization \cite{tibshirani1996regression}, which results in sparse models with several coefficients equal to zero, thus already performing feature selection to some degree. 
For the 30-class isolate identification and 8-class antibiotic treatment classification tasks, we use a multinomial logistic regression model \cite{tutz2015variable}, which outputs a probability distribution across all the classes; the class with the largest estimated probability is taken as the final prediction. 
For the 2-class methicillin resistance classification, standard (binomial) logistic regression is used. 

The logistic regression models are fitted using the \texttt{glmnet} R package~\cite{friedman2010regularization}.
We denote the estimated coefficients for each task as $\hat{\beta}_1(\lambda), \ldots, \hat{\beta}_{2p}(\lambda)$; the parameter $\lambda$ controls the strength of the $\ell_1$ penalty and is tuned by 10-fold cross-validation. \changed{Note that, for the multinomial models, we use an ``ungrouped'' $\ell_1$ penalty, so  that all individual regression tasks within the multinomial model are penalized independently. This leads to the selection of more features overall, although it turns out to yield significantly more accurate predictions for these data compared to the alternative ``grouped'' penalty. See~\cite{friedman2010regularization} for more details on how these models are estimated.}
The $\hat{\beta}_j(\lambda)$ and $\hat{\beta}_{j+p}(\lambda)$ coefficients are used to define a score, namely $W_j = |\hat{\beta}_{j}(\lambda)|-|\hat{\beta}_{j+p}(\lambda)|$, for each of the $p$ original features, 
as explained further in Appendix~\ref{app-methods}.
Feature selection is then performed by selecting variables with $W_j \geq T$, 
where $T$ is a data-adaptive threshold computed by the knockoff filter \cite{barber2015controlling}
to control the FDR below 10\%, so that we can expect about 90\% of the
selected features to be important or redundant \cite{candes2018}. 
We denote by $\hat{S} \subseteq \{1,\ldots,p\}$, or $\hat{S}'$, the subset of features thus selected from $X$, or $X'$, respectively.

\subsection{Classification}

\begin{figure}[!htb]
\centering
\includegraphics[width=0.5\textwidth]{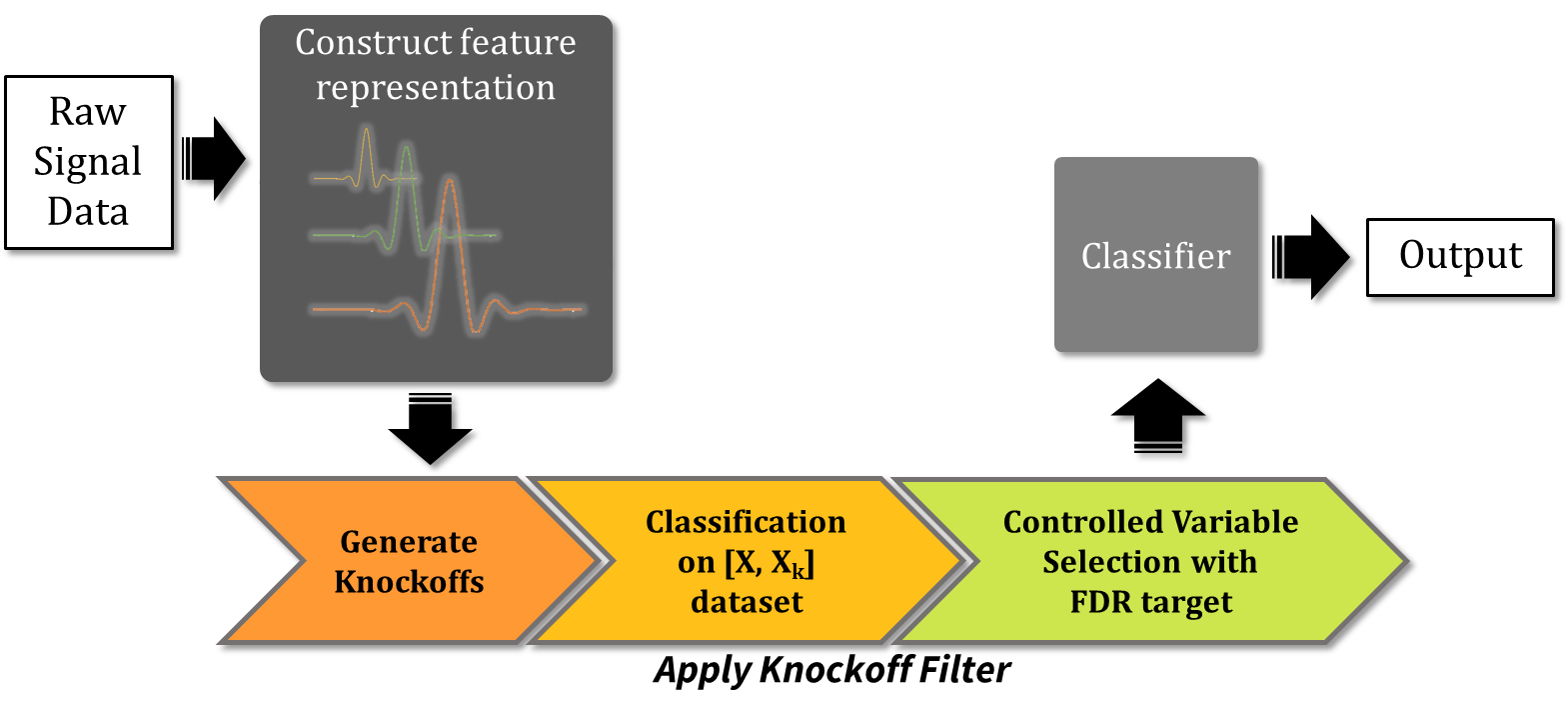}
\caption{Analysis framework for \changed{the interpretable classification of} bacterial Raman spectra. First, wavelet features are extracted from the raw data; then, knockoffs are used to select a predictive and non-redundant subset of them; finally, a \changed{logistic regression} model is fitted on the selected features.}
\label{fig:framework}
\end{figure}

We compare the predictive performance of the features selected by the knockoff filter, $\{X_j\}_{j \in \hat{S}}, \{X'_j\}_{j \in \hat{S}'}$, to that of the full sets of raw and wavelet features, $X,X'$.
For this purpose, we \changed{fit} classification models based on $\ell_1$-regularized (multinomial) logistic regression \changed{on the training data}, as before, for all 12 prediction tasks arising from combination of these four input data sets and the three sets of output labels: 
\begin{equation}
  \begin{bmatrix}
    X \\
    \{X_j\}_{j \in \hat{S}} \\
    X' \\
    \{X'_j\}_{j \in \hat{S}'}
  \end{bmatrix} 
\times
  \begin{bmatrix}
    Y_{30-\text{class}} \\
    Y_{8-\text{class}} \\
    Y_{2-\text{class}}
  \end{bmatrix}.
\end{equation}
Figure~\ref{fig:framework} summarizes \changed{the main steps of our analysis}.
\changed{The out-of-sample predictive performance is evaluated on the 20\% of observations assigned to the test set. The entire analysis is repeated five times, starting from the feature selection step, so that each sample in the full data set is assigned to the test set exactly once. This approach reduces the variability of our findings and facilitates the comparison with the benchmarks from~\cite{ho2019rapidPreprint}.
}
Looking at the results from these 12 prediction tasks, we can evaluate both (A) the effect of applying feature extraction, and (B) the effect of feature selection via the knockoff filter.
Finally, we compare the performance of our models to previous results in \cite{ho2019rapidPreprint}, which were obtained using a CNN, a support vector machine (SVM), and logistic regression models based on different features. 

\section{RESULTS AND DISCUSSION}

Table~\ref{tab:performance-extraction} summarizes the prediction errors obtained by our method for the three prediction tasks (30, 8, and 2 classes), using each of the four input data sets: $X$, $\{X_j\}_{j \in \hat{S}}$, $X'$, $\{X'_j\}_{j \in \hat{S}'}$. \changed{The prediction errors are defined as the misclassification rates, {\em i.e.}, the proportions of data points with incorrectly predicted labels. These results are averaged over the five disjoint test sets that we consider.}
The rows corresponding to results involving the knockoff filter are shaded and arranged below those corresponding to results obtained without controlled variable selection. 
The third column counts the number of non-zero coefficients in the final regularized logistic regression model, which is fitted on the input features after tuning the parameter $\lambda$ by 10-fold cross-validation.

\begin{table}
\caption{\changed{Performances of logistic regression models based on raw and wavelet features, before and after feature selection. These results are averaged over five random test sets (with standard deviations in parenthesis).}}
\label{tab:performance-extraction}
\centering
\begin{tabular}{cccc}
\toprule
Input & Input features & Nonzero coefficients & Test error (\%)\\
\midrule
\addlinespace[0.3em]
\multicolumn{4}{l}{\textbf{30 classes}}\\
\hspace{1em}$X$ & 992 (0) & 992 (0) & 7.4 (0.2)\\
\hspace{1em}$\{X_j\}_{j \in \hat{S}}$ & 980 (2) & 980 (2) & 7.4 (0.2)\\
\hspace{1em}$X'$ & 1105 (0) & 1018 (8) & 6.8 (0.2)\\
  \rowcolor{gray!20}
\hspace{1em}$\{X'_j\}_{j \in \hat{S}'}$ & 136 (3) & 136 (3) & 5.1 (0.2)\\
\addlinespace[0.3em]
\multicolumn{4}{l}{\textbf{8 classes}}\\
\hspace{1em}$X$ & 992 (0) & 991 (1) & 5.6 (0.1)\\
\hspace{1em}$\{X_j\}_{j \in \hat{S}}$ & 949 (4) & 949 (4) & 5.5 (0.2)\\
\hspace{1em}$X'$ & 1105 (0) & 1008 (4) & 5.3 (0.2)\\
  \rowcolor{gray!20}
\hspace{1em}$\{X'_j\}_{j \in \hat{S}'}$ & 111 (8) & 111 (8) & 4.7 (0.3)\\
\addlinespace[0.3em]
\multicolumn{4}{l}{\textbf{2 classes}}\\
\hspace{1em}$X$ & 992 (0) & 668 (36) & 7.3 (0.6)\\
\hspace{1em}$\{X_j\}_{j \in \hat{S}}$ & 478 (38) & 477 (37) & 7.5 (0.6)\\
\hspace{1em}$X'$ & 1105 (0) & 254 (107) & 6.1 (0.4)\\
  \rowcolor{gray!20}
\hspace{1em}$\{X'_j\}_{j \in \hat{S}'}$ & 66 (8) & 66 (8) & 6.2 (0.3)\\
\bottomrule
\end{tabular}
\end{table}

\subsection{Effect of feature extraction}

To examine the effect of feature extraction, we compare the classification errors in Table~\ref{tab:performance-extraction} corresponding to the input data sets $X$ and $X'$ (in the white rows), for each of the three tasks. In each case, we observe a decrease in test error, from \changed{7.4\% to 6.8\%} for the 30-class task, \changed{from 5.6\% to 5.3\%} for the 8-class task, and from \changed{7.3\% to 6.1\%} for the 2-class task. 

As an additional comparison, Table~\ref{tab:performance-selection} reports the predictive performance (within a regularized logistic regression model) of our wavelet features next to that of features obtained by performing a component analysis (PCA) on the raw signal data. (Recall that PCA extracts directions with maximal variance in the data matrix.)
To facilitate the comparison, the number of principal components is fixed to match the number of wavelet features selected by the knockoff filter for each classification task.
Again, the wavelet features yield lower classification errors, which should not be very surprising given that they have a much more intuitive interpretation for our kind of data.

\begin{table}
\caption{Performances of logistic regression models based on raw-PCA and knockoff-filtered wavelet features. \changed{Other details are as in Table~\ref{tab:performance-extraction}.}}
\label{tab:performance-selection}
\centering
\begin{tabular}{cccc}
\toprule
Input & Input features & Nonzero coefficients & Test error (\%)\\
\midrule
\addlinespace[0.3em]
\multicolumn{4}{l}{\textbf{30 classes}}\\
\hspace{1em}$X_{\text{PCA}136}$ & 136 (3) & 132 (3) & 5.4 (0.2)\\
  \rowcolor{gray!20}
\hspace{1em}$\{X'_j\}_{j \in \hat{S}'}$ & 136 (3) & 136 (3) & 5.1 (0.2)\\
\addlinespace[0.3em]
\multicolumn{4}{l}{\textbf{8 classes}}\\
\hspace{1em}$X_{\text{PCA}111}$ & 111 (8) & 98 (7) & 5.2 (0.2)\\
  \rowcolor{gray!20}
\hspace{1em}$\{X'_j\}_{j \in \hat{S}'}$ & 111 (8) & 111 (8) & 4.7 (0.3)\\
\addlinespace[0.3em]
\multicolumn{4}{l}{\textbf{2 classes}}\\
\hspace{1em}$X_{\text{PCA}66}$ & 66 (8) & 42 (5) & 7.8 (0.2)\\
  \rowcolor{gray!20}
\hspace{1em}$\{X'_j\}_{j \in \hat{S}'}$ & 66 (8) & 66 (8) & 6.2 (0.3)\\
\bottomrule
\end{tabular}
\end{table}

\subsection{Effect of feature selection}

To examine the effect of feature selection with the knockoff filter, we compare the classification errors in adjacent pairs of rows in Table~\ref{tab:performance-extraction}, for each of the three tasks. 
\changed{In general, we observe that controlled feature selection with the knockoff filter can simultaneously improve interpretability, since models based on fewer features are easier to explain, as well as  predictive accuracy.
In particular, we see some improvements in classification accuracy between $X'$ and $\{X'_j\}_{j \in \hat{S}'}$,
as the knockoff filter selects a subset of the wavelet features before the final classifier is trained.
The misclassification rate decreases
from \changed{6.8\% to 5.1\%} for the 30-class task, from \changed{5.3\% to 4.7\%} for
the 8-class task, and remains approximately constant \changed{(it increases slightly from 6.1\% to 6.2\%)} for the 2-class task.
These results are notable given that the numbers of features input
into the classifier are reduced significantly: of the original
\changed{1105} wavelet features, we are left with only \changed{136, 111,
and 66} features for the respective tasks, on average. 
By contrast, the lasso model estimated without knockoff filtering assigns non-zero coefficients to as many as 1018 features (for the 30-class problem), which will clearly make the interpretations more challenging. 
We have also observed that a grouped $\ell_1$ penalty, instead of the ungrouped one we adopted here, would result in a lasso model with fewer variables, but at the cost of lower accuracy.
}
The benefits of controlled feature selection are less obvious when this is applied
to the raw signal data, $X$.
In this case, \changed{the knockoff filter does not reduce the number of features significantly, and the resulting changes in classification errors are minimal.}
This suggests the wavelet features are much more informative, as fewer of them can achieve equivalent or possibly higher accuracy compared to the raw signal variables.

\subsection{Comparisons with other classifiers}
 
Table~\ref{tab:performance-benchmarks} compares the performance of our method to that of other classifiers applied to the same data. 
These benchmarks are a convolutional neural network (CNN), a support vector machine (SVM), and logistic regression (LR) without regularization \cite{ho2019rapidPreprint}. 
The CNN is applied directly to the raw signals, while the SVM and LR take the top 20 principal components as input features \cite{ho2019rapidPreprint}.
We denote our method as KWLR (knockoff-filtered wavelet logistic regression).
Again, \changed{we report the average performance over five disjoint test sets, each containing 20\% of the data points.}

\begin{table}
\caption{\changed{Comparison} of \changed{average} test errors obtained with our method to \changed{those corresponding to} other
models. The results in the white rows are quoted from \cite{ho2019rapidPreprint}.}
\label{tab:performance-benchmarks}
\begin{center}
\begin{tabular}{
  >{\centering\arraybackslash}p{1.0cm}
  >{\centering\arraybackslash}p{1.0cm}
  >{\centering\arraybackslash}p{1.0cm}
  >{\centering\arraybackslash}p{1.3cm}
  >{\centering\arraybackslash}p{0.9cm}
  >{\centering\arraybackslash}p{0.9cm}
  }
  \toprule
    \multirow{3}{*}{Input} & \multirow{3}{*}{Classifier} & \# of input features & \# of nonzero coefficients & Test error (\%) & Error s.~d. (\%) \\
  \midrule
  \multicolumn{2}{l}{\textbf{30 classes}}\\
  $X$  & CNN & 992 & 992 & 6.2 & 0.1 \\
  $X_{\text{PCA20}}$  & SVM & 20 & 20 & 11.3 & 0.2 \\
  $X_{\text{PCA20}}$  & LR & 20 & 20 & 10.7 & 0.2 \\
  \rowcolor{gray!20}
  $\{X'\}_{j \in \hat{S}'}$  & KWLR & 136 & 136 & 5.1 & 0.2 \\

  \addlinespace[0.5em]
  \multicolumn{2}{l}{\textbf{8 classes}}\\

  $X$  & CNN & 992 & 992 & 1.0 & 0.1 \\
  \rowcolor{gray!20}
  $\{X'\}_{j \in \hat{S}}$ & KWLR & 111 & 111 & 4.7 & 0.3 \\

  \addlinespace[0.5em]
  \multicolumn{2}{l}{\textbf{2 classes}}\\

  $X$  & CNN & 992 & 992 & 4.6 & 0.5 \\
  \rowcolor{gray!20}
  $\{X'\}_{j \in \hat{S}'}$  & KWLR & 66 & 66 & 6.2 & 0.3 \\

  \bottomrule
  \end{tabular}
\end{center}
\end{table}

Our proposed method (KWLR) \changed{uses far fewer features compared to the benchmarks; such parsimony, combined with the intuitive nature of the logistic regression model, makes our approach much more easily interpretable.
Furthermore, KWLR leads to more accurate predictions} for the 30-class task, 
which is the most difficult one,
yielding a \changed{misclassification rate} that is almost half that of the SVM and LR. 
\changed{The KWLR predictions for the 8-class and 2-class prediction tasks are less accurate than those obtained with the CNN.
This result may be explained by noting that the signal-to-noise ratio is higher compared to the 30-class classification problem, 
as there are now more examples per class and clearer distinctions between them. 
In this low-noise setting, the higher flexibility of the CNN provides the latter with an advantage, without necessarily also
involving higher risk of overfitting.
In any case, our method achieves the primary goal set by this paper: we improve interpretability, as our model is simpler and utilizes far fewer features compared to the CNN, while achieving satisfactory predictive accuracy.

Table~\ref{tab:performance-extraction-others} in Appendix~\ref{app-results} compares the performances of Naive Bayes and Nearest-Neighbor classification models based on our wavelet features, before and after feature selection with knockoffs. These alternatives are simple and easy to interpret, but not as accurate as logistic regression. Nonetheless, feature selection with knockoffs still tends to relatively improve the predictive accuracy of the Naive Bayes and Nearest-Neighbor models, while also greatly reducing the number of features upon which their output depends. In conclusion, our results suggest logistic regression with the wavelet features selected by the knockoff filter achieves the best trade-off between interpretability and predictive accuracy for this data set.
}

\subsection{Visualization of the wavelet features}

\begin{figure}[!htb]
\centering
\includegraphics[width=0.3\textwidth]{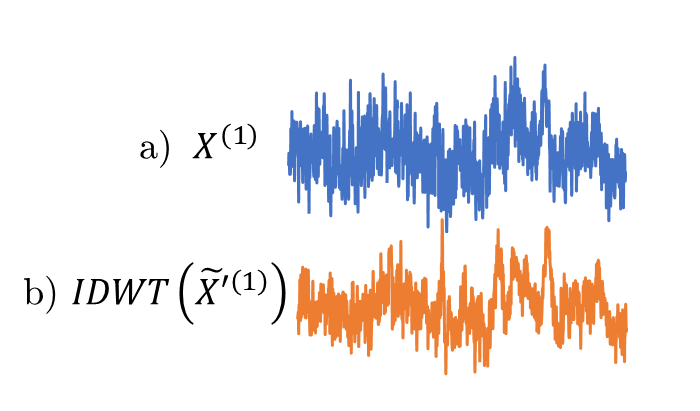}
\caption{(a) Raman signal, and (b) knockoff copy
of its wavelet representation, projected back into
the signal domain.
}
\label{fig:knockoffs-visualize}
\end{figure}

Figure~\ref{fig:knockoffs-visualize}(a) shows an example of a raw Raman signal, which we denote as $X^{(1)}$. 
From this, we extract wavelet features $X'^{(1)}$ and generate corresponding knockoffs $\tilde{X}'^{(1)}$.
Figure~\ref{fig:knockoffs-visualize}(b) shows the IDWT projection of $\tilde{X}'^{(1)}$ back into the signal domain.
We observe that the knockoff signal preserves some characteristics of the original signal, such as its general shape and noise pattern, but it is clearly distinct. 

\begin{figure}[!htb]
\centering
\includegraphics[width=0.35\textwidth]{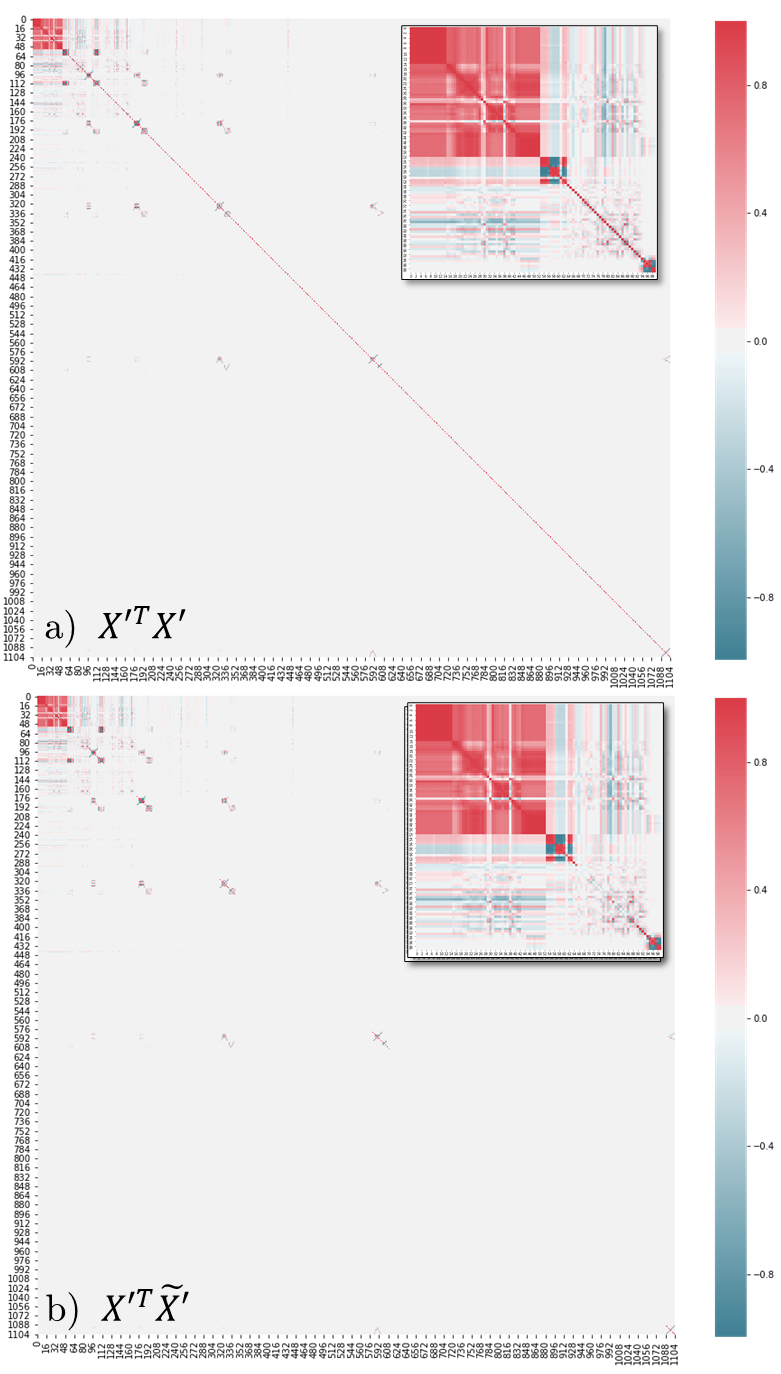}
\caption{Correlation map for (a) original wavelet features; (b)
original and knockoff wavelet features. 
The inset highlights the first 100 features.
}
\label{fig:knockoffs-correlation}
\end{figure}

Figure~\ref{fig:knockoffs-correlation} plots the correlations between the original features (a) and the cross-correlations between the original and knockoff wavelet features (b). 
The first 100
features or so, corresponding to the lower level DWT
coefficients, show the strongest local (among adjacent features) cross-correlations (see insets), while most other features are
approximately uncorrelated.
The property in~\eqref{eq:covariance}, which follows from the construction of the knockoffs, implies that Figure~\ref{fig:knockoffs-correlation}(b) should look very similar to Figure~\ref{fig:knockoffs-correlation}(a), 
except for the values on the diagonal, which can be lower in (b).
This suppression of the diagonal values reflects our attempt to make the knockoffs as different as possible from the 
real features \cite{candes2018} (this can only be partly achieved for the first 100 wavelets because they have stronger correlations among themselves).

This second-order construction of knockoffs assumes a multivariate Gaussian approximation for the
feature distribution, which may not necessarily be very accurate. 
In any case, our classification results indicate that the second-order
knockoffs are effective in performing controlled feature
selection, while retaining power in the selected features.
\changed{The alternative knockoff construction described in \cite{romano2019} can model the 
underlying feature distribution more flexibly, which can make the FDR control more robust if the features are non-Gaussian
but did not make a significant difference in this case.} 

\begin{figure}[!htb]
\centering
\includegraphics[width=0.5\textwidth]{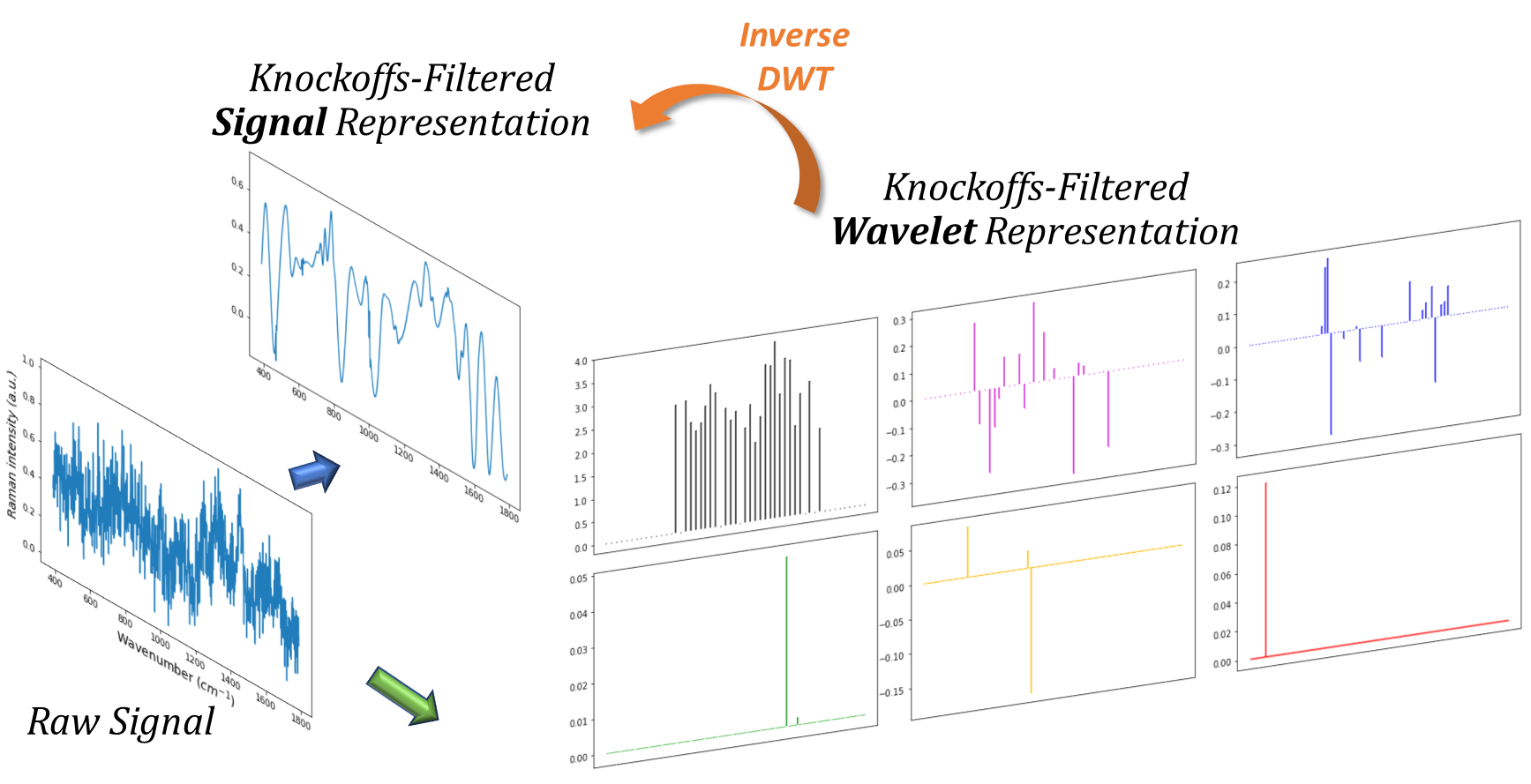}
\caption{Visualization in the wavelet and signal domain of features selected by the knockoff filter.}
\label{fig:selected-features}
\end{figure}

Figure~\ref{fig:selected-features} visualizes a set of knockoff-filtered features in the wavelet and signal domains, with the latter obtained through an IDWT.
Most of these wavelets are at lower frequency, as the higher frequency one tend to be filtered out.
Thus, most noise in the signal domain is removed, while certain peaks are accentuated.
Such peaks reveal interpretable structures that are important for bacterial classification, in a way that the noisy raw signal cannot directly capture.
In particular, we expect peaks in our Raman spectra to indicate distinguishing chemical bonds
found in different classes of bacteria.

\begin{figure}[!htb]
\centering
\includegraphics[width=0.5\textwidth]{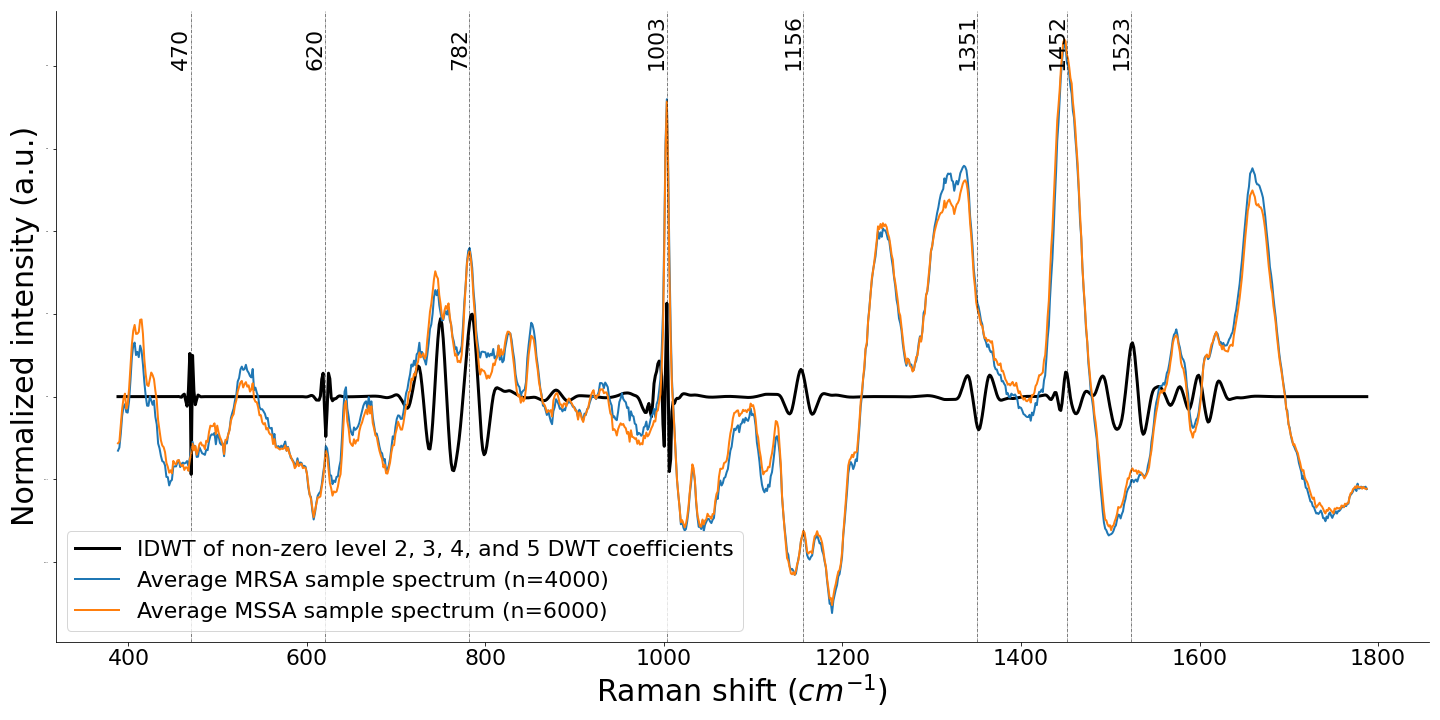}
\caption{Visualization in the signal domain of level 2, 3, 4, and 5 wavelet features selected by the knockoff filter for the 2-class (MRSA vs. MSSA) prediction task (black), along with the averaged spectra from each class (color). Many of the selected features have peaks (vertical lines) at wavelengths corresponding to known relevant chemical bonds.}
\label{fig:nonzero-features}
\end{figure}

Examining the higher order ({\em i.e.}, more spatially localized) wavelet features selected by our method for the 2-class task, we indeed observe these are consistent with peaks previously identified as being relevant to discriminating between methicillin-resistant (MRSA) and methicillin-sensitive strains (MSSA) of the Staphylococcus aureus bacteria \cite{ayala2018drug}. Recall that we say a feature is ``selected'' if its fitted coefficient is non-zero after performing knockoff filtering and sparse logistic regression modeling. The non-zero detail coefficients from levels 2, 3, 4, and 5 of the wavelet transform for a single MRSA sample are represented in Figure~\ref{fig:selected-features} by spikes in the blue, green, yellow, and red plots, respectively. Figure~\ref{fig:nonzero-features} shows the IDWT of these features alongside the averaged spectra from each class. Specifically, it appears that level 2 wavelets with peaks close to 781 cm$^{-1}$, 1004 cm$^{-1}$, 1159 cm$^{-1}$, and 1523 cm$^{-1}$ were selected. These correspond to the breathing modes for the pyrimidine ring and phenylalanine, as well as the C-C and C=C stretching modes for staphyloxanthin, a carotenoid pigment produced by \emph{S. aureus} that gives it its characteristic golden color. Further, level 3 and 4 wavelets with peaks around 1456 cm$^{-1}$ and 1004 cm$^{-1}$ were also selected, pointing to the CH$_{2}$/CH$_{3}$ bending mode and phenylalanine breathing mode. These findings agree with the considerably less noisy Raman microspectroscopy data in \cite{ayala2018drug}, which found that the ratios of the 1159 cm$^{-1}$, 1523 cm$^{-1}$, and 1456 cm$^{-1}$ peaks to the 1004 cm$^{-1}$ peak were highly predictive of methicillin resistance, and potentially also indicate differences in pigmentation and lipid concentration in \emph{S. aureus} strains. In addition to these known peaks, we also observed that level 5, 4, and 2 wavelets corresponding respectively to peaks around 470 cm$^{-1}$, 620 cm$^{-1}$, and 1351 cm$^{-1}$ were important. It is possible that future research will shed light on their chemical origins and allow us to more fully understand the phenotypic differences between methicillin-resistant and methicillin-sensitive \emph{S. aureus} bacteria. 

\section{CONCLUSION}

This paper demonstrates that a combination of feature extraction guided by domain knowledge and 
controlled variable selection via knockoffs can improve model interpretability 
for learning tasks involving signal data, 
relative to more obscure machine learning algorithms.
\changed{At the same time, this may even improve predictive accuracy.
Whether the proposed approach should be preferred in practice to black-box machine learning depends on the data at hand,
as well as on the importance given by practitioners to interpretability.
The bacterial classification problem considered in this paper is one example in which interpretability is particularly desirable, 
especially if the final predictions will be reviewed by an expert who needs to understand how the model reaches its conclusions in order to make the best informed decision.
}

The approach described in this paper also has the advantage of being
computationally more affordable to train compared to many typical black-box ML algorithms,
as we utilize \changed{models with fewer parameters and input features}. 
\changed{Although a particularly simple logistic regression classifier worked well for our application, 
our method can easily accommodate more flexible models, 
possibly capturing nonlinear relations and thus leading to even more accurate predictions.}

In conclusion, \changed{we have presented an example of} a systematic and principled approach to the
analysis of signal data, which can facilitate the development of human-interpretable 
models with good predictive performance.
Future research may explore the use of more automated feature
extraction models within our framework, such DeepPINK \cite{lu2018deeppink},
or more complex learning algorithms accompanied by an
appropriate quantitative measure of feature importance, {\em e.g.},
SHAP values \cite{lundberg2017unified}. 
Alternative knockoff generation algorithms
such as that in \cite{romano2019} could enhance the robustness of the feature selection step and thus
further improve predictive accuracy. Finally, it
would be interesting to investigate the impact of the
FDR level on the predictive accuracy of our \changed{method}. In this
paper, we have focused on the standard level of 10\% for
simplicity, and because larger values did not seem to bring
much improvement\changed{; see Figure~\ref{fig:knockoffs-fdr} in Appendix~\ref{app-results}}. However, the optimal choice may
generally be data-dependent.

\printbibliography

\begin{appendix}

\subsection{Review of the knockoff filter method} \label{app-methods}

\changed{Here, we recall} the knockoff framework \cite{candes2018}. 
We consider $n$ independent pairs of observations $(X^{(i)}, Y^{(i)})$, such that $Y^{(i)}$ depends on $X^{(i)} = (X_1^{(i)}, \ldots, X_p^{(i)})$, with $i \in \{1, \ldots, n\}$, through some unknown conditional distribution $F_{Y \mid X}$:
\begin{equation*}
  Y^{(i)} \mid X_1^{(i)}, \ldots, X_p^{(i)} \sim F_{Y \mid X}.
\end{equation*}
We seek to find the smallest subset of {\em important} features,
$S \subseteq \{1,\ldots,p\}$, upon which $F_{Y \mid X}$ depends; 
{\em i.e.}, $Y$ should be independent of $\{X_j\}_{j \not\in S}$ conditional on $\{X_j\}_{j \in S}$.
We denote by $H_0 = \{1,\ldots,p\} \setminus S$ the set of {\em null} (unimportant) features.

The FDR for some $\hat{S} \subseteq \{1,\ldots,p\}$
is defined as the expected fraction of null features among it:
\begin{equation*}
  \text{FDR} = \mathbb{E} \left[ \frac{|\hat{S} \cap H_0|}{\max(1, |\hat{S}|)} \right].
\end{equation*}

The goal is to discover as many
important features as possible while keeping the FDR below a
specified level. 
This can be achieved by generating, in silico, a {\em knockoff} copy $\tilde{X}$ of $X$, which 
should satisfy the following two properties:
\begin{enumerate}
  \item $Y$ is independent of $\tilde{X} \mid X$;
  \item $[X,\tilde{X}]$ and $[X,\tilde{X}]_{\text{swap}(j)}$ have the same distribution, for any $j \in \{1,\ldots,p\}$, 
where $[X,\tilde{X}]_{\text{swap}(j)}$ is the vector obtained by swapping $X_j$ with $\tilde{X}_j$.
\end{enumerate}
The first condition above simply states that knockoffs are null; this is immediately guaranteed if $\tilde{X}$ is generated before looking at $Y$.
The second condition states that the features in $X$ and $\tilde{X}$ are
pairwise exchangeable, which implies that null features have
on average the same explanatory power for $Y$ as their corresponding
knockoffs.
These two properties allow knockoffs to serve as negative controls \cite{candes2018}, as explained below.
Note that the equality in distribution (second property) is generally difficult to enforce exactly, so we make
some approximation and only match the first two moments,
following in the footsteps of the previous literature \cite{candes2018,romano2019}.

After augmenting the original feature matrix with the knockoffs, {\em i.e.}, as $[X, \tilde{X}] \in \mathbb{R}^{n \times 2p}$,
a learning model is trained to predict $Y$, from which feature importance measures $Z=(Z_1, \ldots, Z_{2p})$ are then extracted for each of the augmented features.
For example, if we adopt a simple regularized logistic regression model, we can define the scores $Z_j = |\hat{\beta}_j(\lambda)|$, where $\hat{\beta}(\lambda)$ denotes the vector of estimated coefficients, and $\lambda$ is the regularization parameter tuned by cross-validation. 
\changed{In the case of the 30-class and 8-class problems, we apply a multinomial logistic regression model and define the scores $Z_j$ and $\tilde{Z}_j$ as the sum of the absolute regression coefficients obtained for each regression task.} 
Ideally, we would always like null features to have $Z_j$ close
to zero, although this may not generally be the case in practice;
hence the need to calibrate these measures through the knockoffs. 

For each $j \in \{1,\ldots,p\}$, an importance statistic $W_j$ is defined by contrasting $Z_j$ with $Z_{j+p}$, 
{\em i.e.}, $W_j = Z_j -Z_{j+p}$.
Therefore, $W_j > 0$ indicates that $X_j$ appears to be more important
than its knockoff copy, which provides evidence against the null
hypothesis $j \in H_0$. 

By construction, each $W_j$ has equal probability of being
positive or negative if $j \in H_0$; more precisely, the signs of null $W_j$ are
independent and identically distributed flips of a fair coin \cite{barber2015controlling}.
The knockoff filter leverages this property to select features with sufficiently large $W_j$,
according to a data-adaptive threshold that depends on the desired FDR level.
The intuition is that the proportion of false discoveries in $\{j : W_j \geq t\}$ can be estimated conservatively by:
\begin{equation*}
  \widehat{\text{FDP}}(t) = \frac{1+| \{ j : W_j \leq -t \}|}{|\{j : W_j \geq t\}|}.
\end{equation*}
In particular, the FDR can be provably controlled \cite{barber2015controlling} below $q \in (0,1)$ by selecting 
$\hat{S} = \{j : W_j \geq T\}$, with $T = \min \{ t : \widehat{\text{FDP}}(t) \leq q \}$.

\subsection{\changed{Comparison with other variable selection algorithms}} \label{app-results}

\begin{table}[!htb]
\caption{\changed{PERFORMANCES OF ALTERNATIVE MODELS WITH WAVELET
FEATURES. OTHER DETAILS ARE AS IN TABLE~\ref{tab:performance-extraction}.}}
\label{tab:performance-extraction-others}
\centering
\begin{tabular}{cccc}
\toprule
Method & Input & Input features & Test error (\%)\\
\midrule
\addlinespace[0.3em]
\multicolumn{4}{l}{\textbf{30 classes}}\\
\hspace{1em}Nearest Neighbor & $X'$ & 1105 (0) & 62.8 (1.6)\\
\hspace{1em}Nearest Neighbor & $\{X'_j\}_{j \in \hat{S}'}$ & 16 (0) & 27.7 (0.7)\\
\hspace{1em}Naive Bayes & $X'$ & 1105 (0) & 27.0 (0.5)\\
\hspace{1em}Naive Bayes & $\{X'_j\}_{j \in \hat{S}'}$ & 29 (7) & 35.6 (3.8)\\
\addlinespace[0.3em]
\multicolumn{4}{l}{\textbf{8 classes}}\\
\hspace{1em}Nearest Neighbor & $X'$ & 1105 (0) & 34.9 (2.0)\\
\hspace{1em}Nearest Neighbor & $\{X'_j\}_{j \in \hat{S}'}$ & 10 (0) & 24.9 (0.2)\\
\hspace{1em}Naive Bayes & $X'$ & 1105 (0) & 39.3 (1.1)\\
\hspace{1em}Naive Bayes & $\{X'_j\}_{j \in \hat{S}'}$ & 40 (3) & 33.5 (0.4)\\
\addlinespace[0.3em]
\multicolumn{4}{l}{\textbf{2 classes}}\\
\hspace{1em}Nearest Neighbor & $X'$ & 1105 (0) & 38.9 (0.9)\\
\hspace{1em}Nearest Neighbor & $\{X'_j\}_{j \in \hat{S}'}$ & 14 (3) & 21.1 (3.9)\\
\hspace{1em}Naive Bayes & $X'$ & 1105 (0) & 25.0 (2.4)\\
\hspace{1em}Naive Bayes & $\{X'_j\}_{j \in \hat{S}'}$ & 25 (2) & 22.5 (1.5)\\
\bottomrule
\end{tabular}
\end{table}

\begin{figure}[!htb]
\centering
\includegraphics[width=0.5\textwidth]{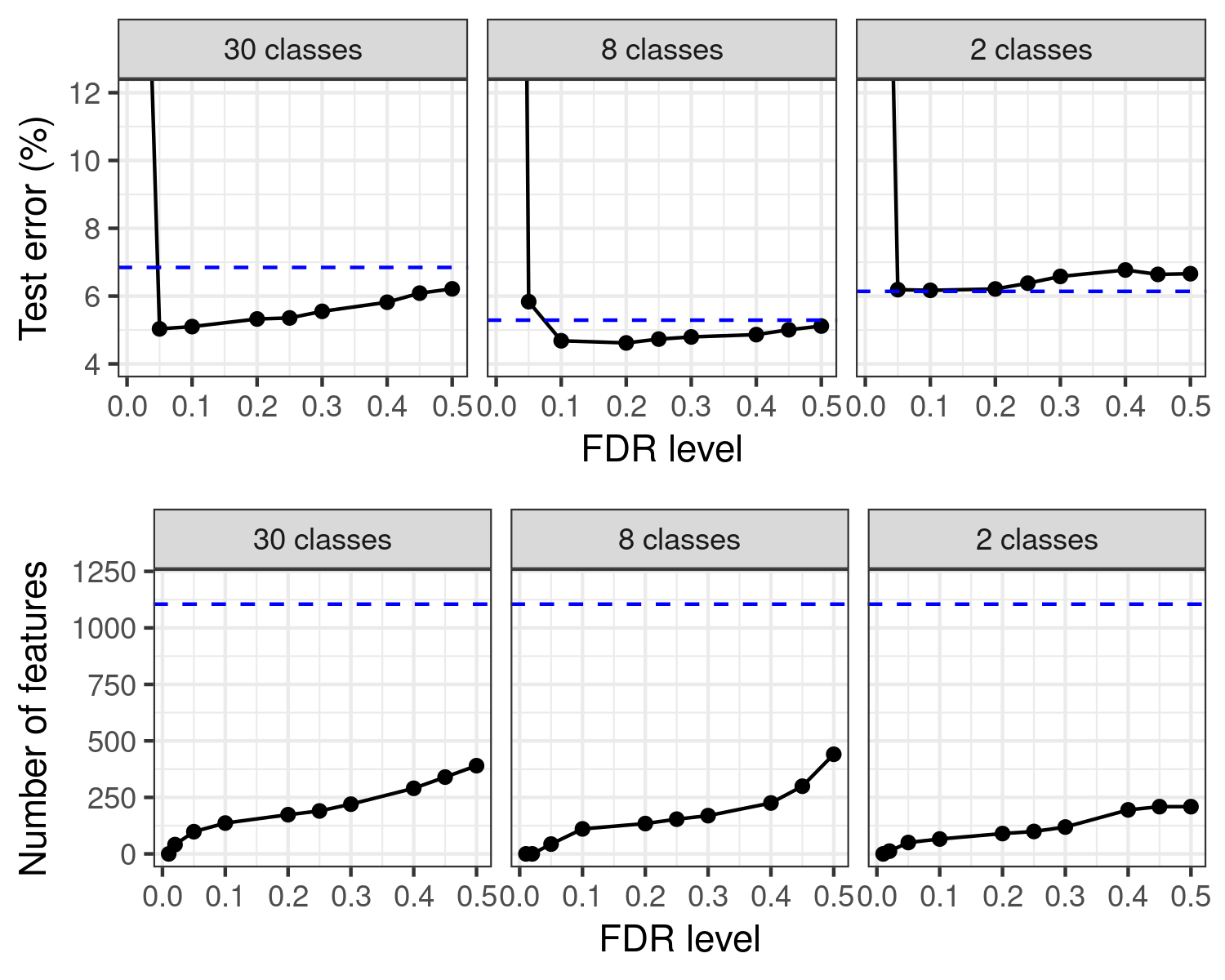}
\caption{\changed{Model performance after selection of wavelet features with knockoffs, using different FDR levels.
The horizontal dashed lines correspond to the results obtained with the lasso tuned by cross-validation, without applying the knockoff filter. 
Other details are as in Table~~\ref{tab:performance-extraction}.}}
\label{fig:knockoffs-fdr}
\end{figure}

\end{appendix}

\end{document}